\documentclass[twocolumn,times]{aastex62}
\usepackage{amsmath}
\usepackage{amssymb}
\usepackage{lineno}
\usepackage{bm}
\allowdisplaybreaks[4]

\def\be {\begin{equation}}
\def\ee {\end{equation}}

\def\tt {\{t\}}

\usepackage{color}

\shorttitle{Origin of the Moon}
\shortauthors{Wenshuai Liu}
\begin{document}

%\title{Slow radial migration of a gap-opening planet triggered by dust feedback}
\title{\large{\textbf{Origin of the lunar farside highlands from Earthshine-induced global circulation in lunar magma ocean}}}

\correspondingauthor{Wenshuai Liu}
\email{674602871@qq.com}

\author{Wenshuai Liu}
\affiliation{School of Physics, Henan Normal University, Xinxiang 453007, China}



\begin{abstract}
The lunar farside highlands, referred to as the lunar farside thicker crust compared with the nearside crust, presents a challenge to the theory of formation and evolution of the Moon. Here, we show that, after the Moon reached synchronous rotation, Earthshine could induce global circulation in lunar magma ocean due to the imposed surface temperature gradient generated by the hot, post-giant impact Earth. The global circulation, generating downwellings on the farside and a deeper return flow on the nearside, results that magmas flow from the nearside to the farside in the shallow magma ocean while the the direction of flow is opposite in the deep magma ocean. Such flow in the shallow magma ocean would transport anorthositic crystals formed in the nearside to the farside. Furthermore, since the lunar farside is cooler than the nearside, crystallization is much more efficient at the farside, resulting that farside magmas transported from the nearside produce anorthositic crystals rapidly. The theory proposed here may provide a natural way of explaining the origin of the lunar farside highlands and the lunar dichotomy.
\end{abstract}



\section{Introduction}

The giant impact theory is the leading theory for the lunar origin, according to which the Moon formed through accreting material from debris produced by a collision between a planet-sized body Theia and the proto-Earth \citep{1,2}. After lunar formation, the Moon is thought to be in a molten state with a global magma ocean from which crystallization would persist along with the cooling of lunar magma ocean. During solidification, olivine and pyroxene sank into the deep mantle while plagioclase floated to the top to form the anorthositic crust, resulting that the residual dense melts full of incompatible-element materials, like ilmenite-bearing cumulate (IBC) and KREEP, resided under the crust. Later, crystallization of such layer of dense melts would overturn onto the core-mantle boundary, leading to a spherically symmetric layered structure inside the Moon \citep{3,4,5}. However, the Moon shows asymmetry of uncertain origin, such as difference in crustal thickness between the Moon's nearside and farside \citep{6}, referred to as the lunar farside highlands problem. In other words, the farside lunar crust is thicker than that in the nearside, posing a challenge to the current theory of lunar formation and evolution. Besides the lunar farside highlands problem, other lunar asymmetries include topography \citep{7}, concentration of incompatible elements \citep{8} and mare volcanic activity \citep{9}. Thus, origin of lunar dichotomy is key to understanding not only lunar evolution but also the process of solidification.

Theories proposed to account for the origin of the dichotomy between hemispheres can be classified into external events and internal process. External events include the accretion of a companion moon onto the Moon \citep{10}, asymmetric nearside-farside cratering \citep{11} and large impact events forming the Procellarum \citep{12} and the South Pole-Aitken basins \citep{13} while internal processes contains asymmetric crystallization of the magma ocean \citep{14,15}, spatial variations in tidal heating \citep{16}, tilted convection \citep{17} and farside condensation of atmospheric and accreting material \citep{18}.

Similar to the global circulation in magma ocean of lava planet induced by its host star \citep{19}, here, we propose that, due to the proximity of the hot, post-giant impact Earth, global circulation inside the lunar magma ocean would be generated by imposed surface temperature gradient. The global circulation would lead magma to flow from the nearside to the farside in the shallow magma ocean and flow from the farside to the nearside in the deep magma ocean while generating downwellings on the farside and a deeper return flow on the nearside. During cooling of lunar magma ocean, both nearside and farside would crystallize, and plagioclase crystallized in the nearside could be transported to the farside. Furthermore, crystallization of plagioclase can be rapid and much more efficient at the farside than that in the nearside due to the cooler temperature in the farside. Combination of these two processes would lead to a thicker lunar crust in the farside. Due to the presence of the thicker crust in the lunar farside, the residual dense melts full of incompatible-element materials would be squeezed into the near side by the self-gravity of the Moon, leading to the concentration of incompatible elements in the nearside after crystallization.

The proposed theory is given in Section 2 in detail and the discussions are in Section 3.

\begin{figure}
     %\begin{tabular}{cc}
            \includegraphics[width=0.5\textwidth]{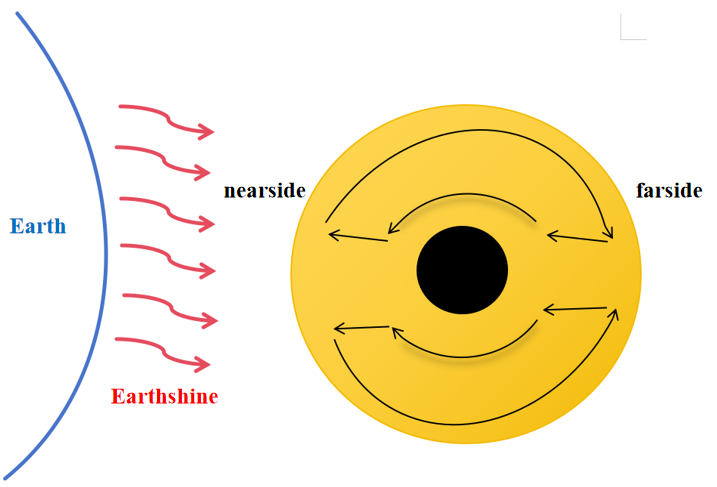}\\
            \includegraphics[width=0.5\textwidth]{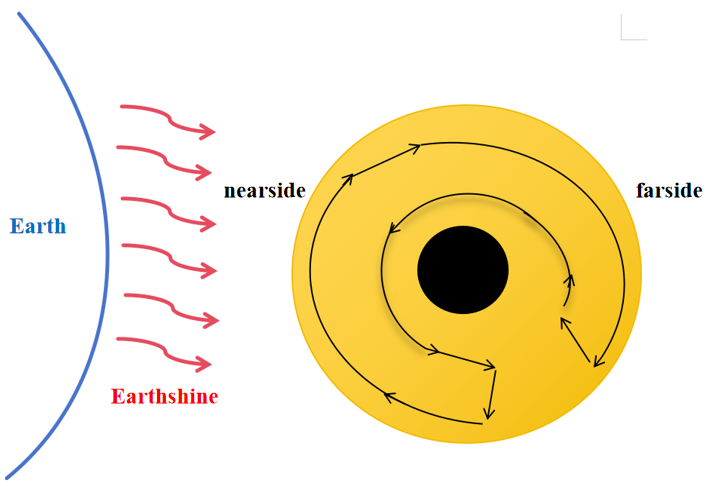}
            %\end{tabular}
\caption{Orange is the lunar magma ocean and black is the lunar core. Global circulation is shown in the lunar magma ocean. Up and bottom show the Moon moving around Earth in counter-clockwise direction with the difference that the Moon in the up has no self-rotation and the Moon in the bottom is in a state of synchronous rotation with the Earth.}
\label{fig:figure1}
\end{figure}

\begin{figure*}
     \begin{tabular}{cc}
            \includegraphics[width=0.33\textwidth]{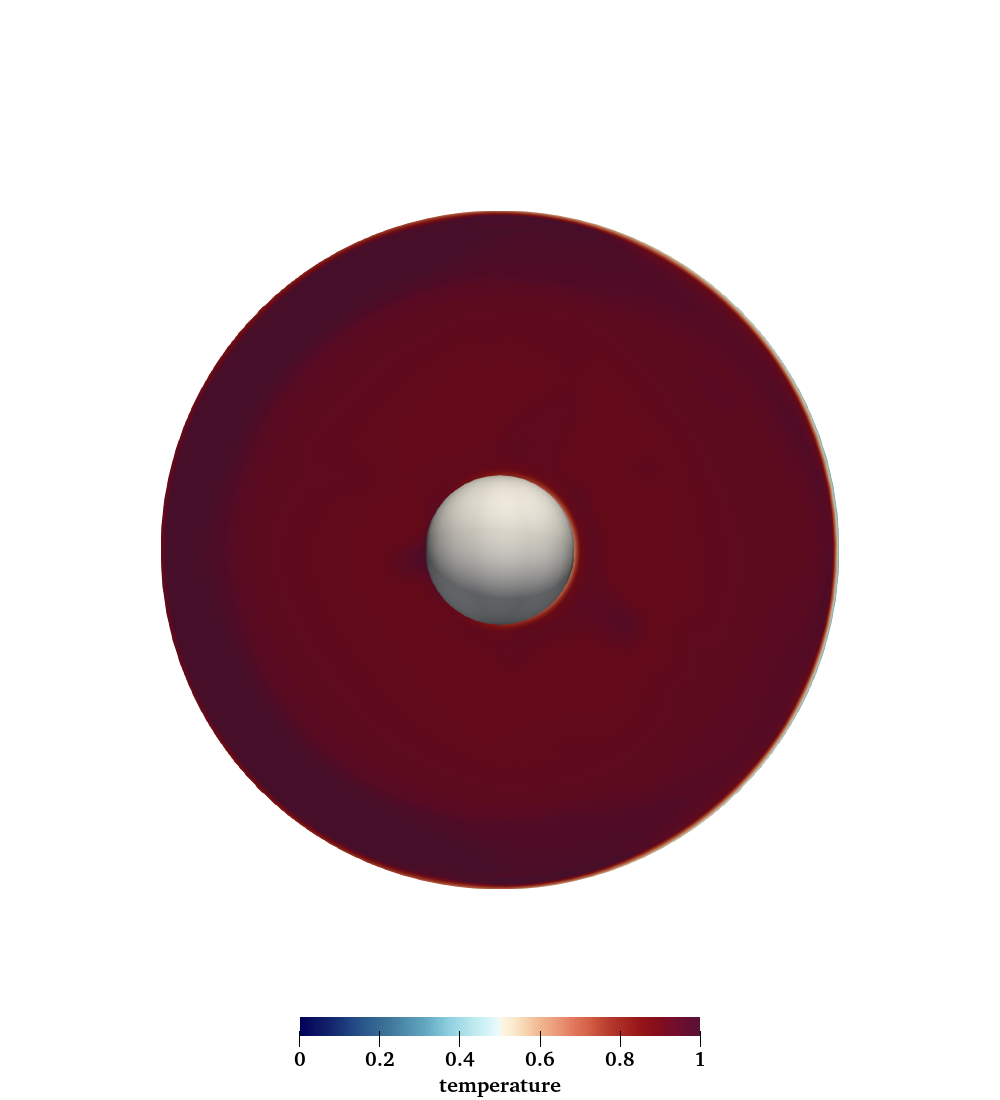}
            \includegraphics[width=0.33\textwidth]{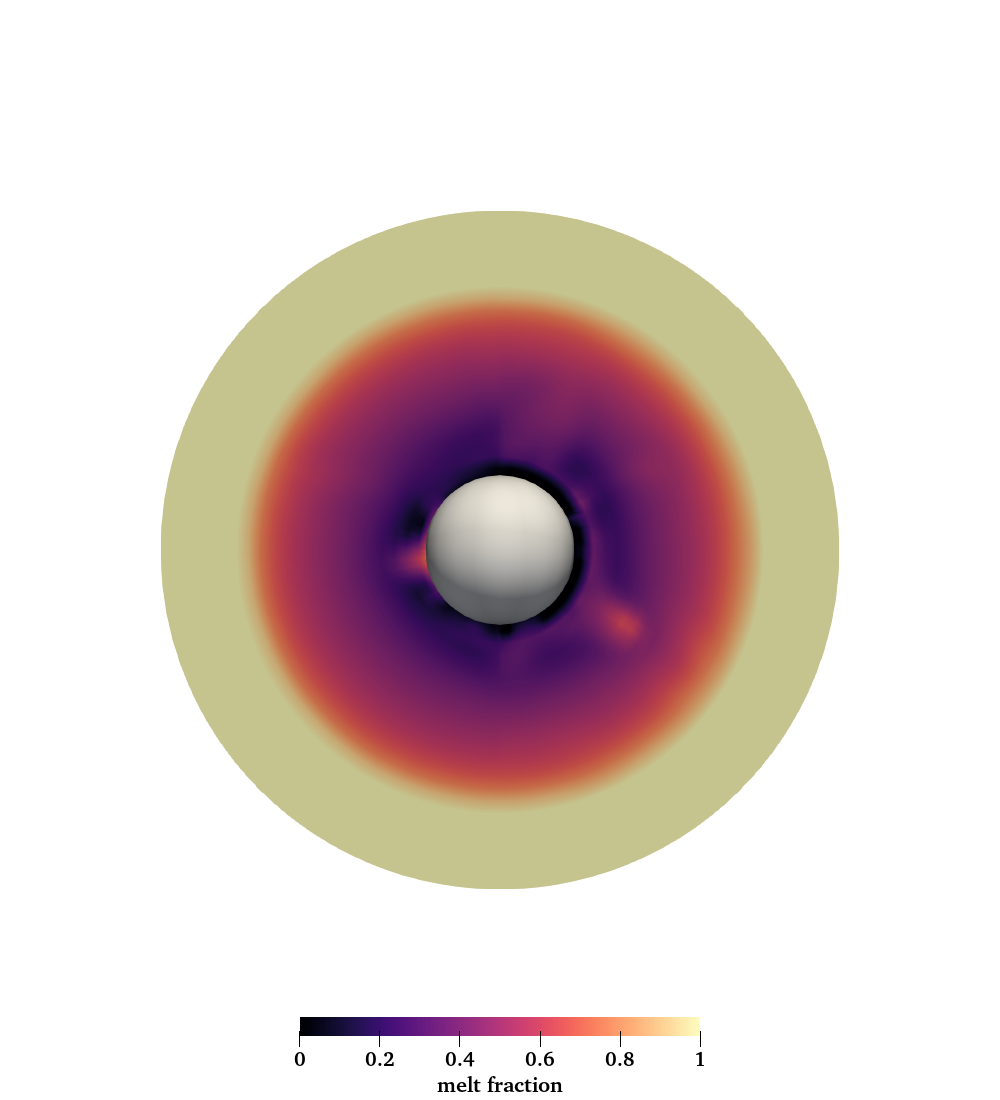}
            \includegraphics[width=0.33\textwidth]{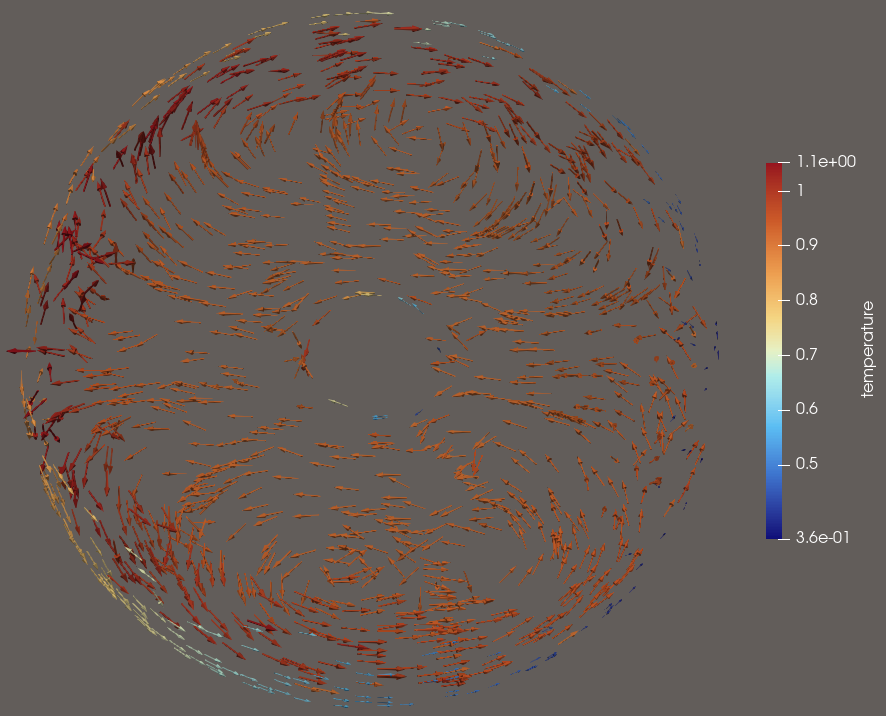}
     \end{tabular}
\caption{Left, middle and right show the 2D cross sections of the temperature, melt fraction and flow patten at model time of 1000 Myr, respectively. Temperature is in unit of 2200K. Earth is on the left of the Moon.}
\label{fig:figure2}
\end{figure*}

\begin{figure}
     %\begin{tabular}{cc}
            \includegraphics[width=0.5\textwidth]{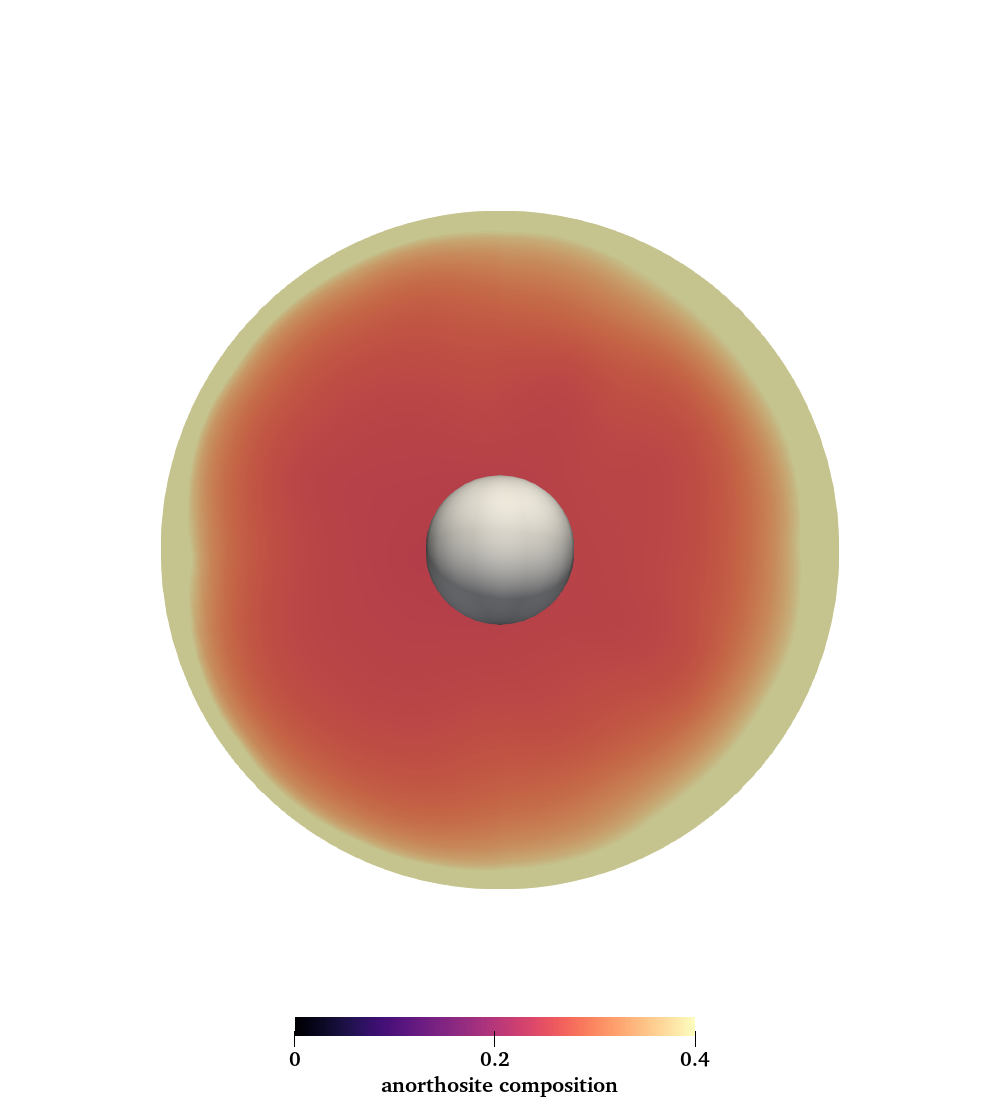}
            %\end{tabular}
\caption{2D cross section of anorthosite composition at model time of 1000 Myr. Earth is on the left of the Moon.}
\label{fig:figure2}
\end{figure}

\begin{figure}
     %\begin{tabular}{cc}
            \centerline{\includegraphics[width=0.35\textwidth]{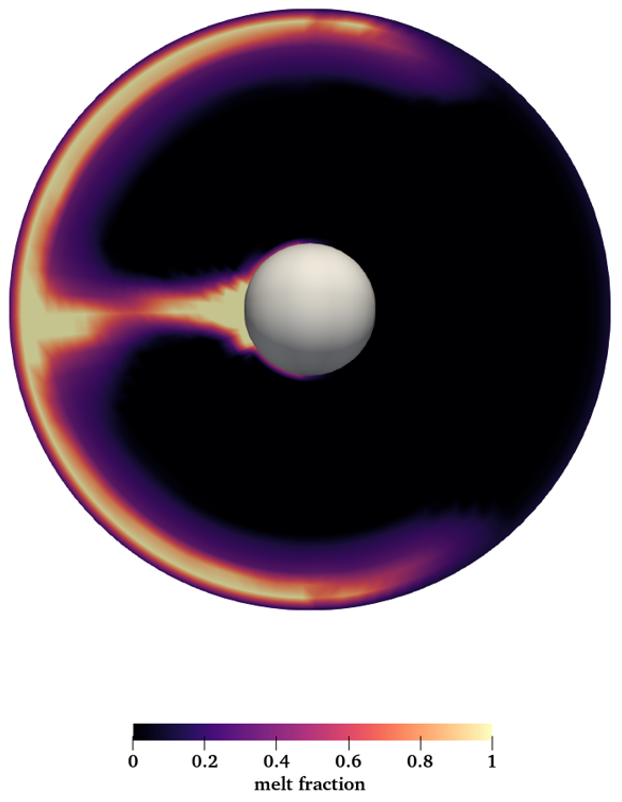}}
            %\end{tabular}
\caption{2D cross section of melt fraction at model time of 1000 Myr with a relatively lower temperature distribution set on the upper boundary and in the simulated region. Earth is on the left of the Moon.}
\label{fig:figure2}
\end{figure}

\section{Global circulation induced by Earthshine}
After the giant impact, circum-terrestrial material produced by the collision between Theia and the proto-Earth is predominantly silicate melt with temperature of 2000K to 5000K exterior to the Roche limit \citep{20}. Since the timescale of lunar formation is much shorter than that of the cooling of the disk to temperatures below the solidus, the Moon would be in a molten state after lunar formation. Afterwards, strong tidal damping inside the Moon would lead the Moon to evolve into a state of synchronous rotation with the Earth quickly, making one side of the Moon always face the Earth during the subsequent evolution. Since the post-giant impact Earth covered by magma ocean is still hot after lunar formation, the effects of Earthshine on the lunar nearside facing Earth would make the nearside much hotter than the farside remaining in the dark, leading to lunar surface temperature asymmetry. It shows from \cite{18} that the lunar farside would cool to an effective temperature of 250K which is set by the solar flux while the nearside could only cool towards $\mathrm{T_{\bigoplus} /2}$ where $\mathrm{T_{\bigoplus}}$ is the temperatures of Earth with $\thicksim 8000$K after the giant impact and could only cool to about $\thicksim 2500$K due to the presence of Earth's atmosphere made of silicate clouds. Thus, the tidally lock molten Moon irradiated by the Earth is similar to the lava planets that orbit so close to their host star that the nearside is hot enough to melt silicate rock.

It shows from \cite{19} that the surface temperature gradient of the lava planet could generate global circulation which drives downwellings on the night-side and a deeper return flow on the day-side and transports magma from the nearside to the farside in the shallow magma ocean and from the farside to the nearside in the deep magma ocean. Thus, global circulation shown in the bottom of Figure 1, similar to that in \cite{19}, would also exist inside the tidally lock molten Moon due to the proximity of the hot, post-giant impact Earth. As lunar magma ocean cooled, dense crystals, such as olivine and pyroxene, sank to lower mantle while buoyant crystal, like plagioclase, floated to the lunar surface, forming the lunar crust.

Due to the presence of global circulation induced by Earthshine, plagioclase crystals formed in the nearside would be transported to the farside along with the magma flow and, during the process of such transportation, the crystallization of new plagioclase crystals would be continue. Finally, the plagioclase crystals formed in the nearside and the newly formed plagioclase crystals during transportation of magma flow from the nearside to the farside would accumulate in the farside. Further more, due to the cooler temperature of the farside, crystallization of plagioclase can be rapid and much more efficient at the farside than that in the nearside, enhancing the amount of plagioclase accumulated in the farside. Thus, the final crust formed in the farside would be thicker than that of the nearside.

In order to confirm the theory we propose above, we conduct simulations with respect to lunar magma ocean crystallization with global circulation induced by Earthshine using CitcomS-Melt \citep{24}. Due to the rapid cooling of the magma ocean's surface of the Moon's farside, a thin anorthositic crust would form on the magma ocean's surface of the Moon's farside. In order to have the same composition at the Moon's nearside as farside, we set bulk mass fraction of anorthite to be $0.98$ and that of olivine to be $0.02$ under the top boundary with depth of 10km. The initial bulk mass fraction of anorthite and olivine in other region is set to be $0.33$ and $0.67$ in order to make the difference of evolved bulk mass fraction of anorthite between the nearside and farside obvious along with time. The temperature of the lower boundary is $\mathrm{T_{lower}=2200K}$ and the initial distribution of temperature inside the simulated region is $\mathrm{\left[\frac{(T_{lower}-T_u)r}{r_{lower}-r_{upper}}+\frac{T_{u}r_{lower}-T_{lower}r_{upper}}{r_{lower}-r_{upper}}\right](1+0.1\sin\theta\cos\phi)}$ where r is the distance to the lunar center, $\mathrm{T_u=1980K}$, $\mathrm{r_{lower}}$ and $\mathrm{r_{upper}}$ are the radius of the lower and upper boundary, respectively. The temperature of the upper boundary is $\mathrm{T_u(0.7+0.3\sin\theta\cos\phi)}$. For simplicity, Coriolis force is neglected in the simualtion. The whole region is divided into 12 caps each of which is divided into $25\times20\times20$ nodes and a grid refinement processing is carried out along the radial direction near the surface (60km thick). Other parameters adopted in CitcomS-Melt are same as that in \cite{24}.

After about 1000 Myr, 2D cross sections of the temperature and melt fraction are shown in the left and middle panel of Figure 2. It shown from Figure 2 that the farside hemisphere is relatively cooler than the nearside hemisphere \citep{25} and solidification of magma ocean appears in the lower region. It should be noted that the fixed temperature distribution with relatively high value set on the nearside of the upper boundary leads to the melt in the upper region with no solidification. In reality, as the Moon recedes from Earth due to tidal evolution, the temperature on the nearside of the upper boundary would decrease and solidification of magma ocean would also appear in the upper region, especially at the farside. The right panel of Figure 2 indicates that magma flows from the nearside to the farside in the shallow magma ocean and flows from the farside to the nearside in the deep magma ocean while generating downwellings on the farside and a deeper return flow on the nearside. Such global circulation would lead to the result presented in Figure 3 which shows that the region of the farside with anorthosite composition larger than 0.4 is thicker than that of the nearside. When the separation between Earth and Moon increases due to tidal evolution, the upper region would solidify, resulting that lunar farside crust would be thicker than the nearside crust and that the farside crust would be older and less evolved than the nearside \citep{14}. Due to the less amount of anorthite and larger amount of olivine in the nearside hemisphere compared with that in the farside hemisphere, the resulting center of mass would be on the left of the geometric center. When Coriolis force, higher resolution and lower temperature of the farside surface are adopted, similar results are obtained.

With a relatively lower temperature distribution set on the upper boundary and in the simulated region compared with that in the simulation described above (meaning a relatively larger separation between Earth and Moon), the resulting 2D cross sections of melt fraction is presented in Figure 4. It shows from Figure 4 that the upper region in the nearside hemisphere would be the place where the residual melts solidify. Furthermore, the area of the residual melts in the nearside shown in Figure 4 would contract along with solidification. Along with crystallization, water tends to reside in the residual melts, residual dense melts are mainly composed of incompatible-element materials, such as ilmenite-bearing cumulate (IBC) and KREEP, meaning that the nearside would be richer in incompatible-element materials and water than farside \citep{26}.

Based on the above, Earthsine would induce asymmetric thermal evolution of the lunar magma ocean and lead to asymmetric crystallization of the lunar magma ocean from the farside to the nearside. From these results, we deduce that Earthsine is the driver of the lunar dichotomy.

\section{Discussions}
We propose a new mechanism for the origin of the lunar farside highlands. After lunar formation, the Moon would be settled into synchronous rotation with the Earth quickly. Due to the proximity of the hot, post-giant impact Earth, lunar surface temperature gradient would be generated by Earthshine. Then, such surface temperature gradient could induce global circulation inside lunar magma ocean. Global circulation of lunar magma ocean induced by Earthshine shows to be a natural way of producing a thicker lunar crust in the farside and, consequently, leading to the lunar dichotomy. The theory proposed here provides a natural way of thermal evolution of lunar magma ocean.

%\bibliographystyle{aasjournal}
%\bibliography{reference} % if your bibtex file is called example.bib

\end{document}